\documentclass{aa}
\usepackage{graphicx}
\usepackage{txfonts}
\usepackage{epsfig}

\begin{document}

\title{Carbon Stars and C/M Ratio in the WLM Dwarf
Irregular Galaxy}

\author{A.~T.~Valcheva\inst{1}
\and
V.~D.~Ivanov\inst{2}
\and
E.~P.~Ovcharov\inst{1}
\and
P.~L.~Nedialkov\inst{1}
}

\offprints{A.~Valcheva}
\institute{Department of Astronomy, University of Sofia,
  5 James Bourchier, Sofia 1164, Bulgaria\\
  \email{valcheva@phys.uni-sofia.bg, evgeni@phys.uni-sofia.bg,
  japet@phys.uni-sofia.bg}
\and  European Southern Observatory, Ave. Alonso de Cordova
  3107, Casilla 19, Santiago 19001, Chile\\
  \email{vivanov@eso.org}
}

\date{Received 18 November 2005 / accepted 18 January 2007}

\authorrunning{Valcheva et al.}

\titlerunning{Near-IR view of WLM: Carbon stars and C/M ratio}

\abstract{ We identify the rich Carbon star population of the
Magellanic-type dwarf irregular galaxy WLM (Wolf-Lundmark-Melotte)
and study its photometric properties from deep near-IR
observations. The galaxy exhibits also a clear presence of Oxygen
rich population. We derive a Carbon to M-star ratio of
C/M\,=0.56$\pm$0.12, relatively high in comparison with many galaxies. 
The spatial distribution of the AGB stars in WLM
hints at the presence of two stellar complexes with a size of a
few hundred parsecs.

Using the H{\sc i} map of WLM and the derived gas-to-dust ratio
for this galaxy N(H{\sc i})/E($B$ $-$ $V$)=60$\pm$10~[10$^{21}$ at.
cm$^{-2}$ mag$^{-1}$] we re-determined the distance modulus of WLM
from the IR photometry of four known Cepheids, obtaining
(m$-$M)$_0$=24.84$\pm$0.14\,mag. In addition, we determine the
scale length of 0.75$\pm$0.14\,kpc of WLM disk in $J$-band.

}
\maketitle

\section{Introduction}

WLM (Wolf-Lundmark-Melotte; also DDO\,221; Wolf \cite{wol09},
Melotte \cite{mel26}) is a dwarf irregular Local Group member.
Earlier photographic surveys of the galaxy were presented in Ables
\& Ables (\cite{abl77}) and Sandage \& Carlson (\cite{san85}).
The first CCD observations of WLM were performed by Ferraro et al.
(\cite{fer89}) who reported significant variations in the recent
star-formation rate across the galaxy. They also detected uniform
underlying relatively old stellar population ($\sim$1\,Gyr). Later on,
Minniti \& Zijlstra (\cite{min97}) derived from $V$ and $I$-band
photometry a distance modulus (m-M)$_0$=24.75$\pm$0.1 mag and
[Fe/H]=\,$-$1.45$\pm$0.2. Hodge et al. (\cite{hod99})
reported the first {\it HST} observations of WLM. They resolved
the sole globular cluster (Ables \& Ables \cite{abl77}) and
obtained (m-M)$_0$=24.73$\pm$0.07 mag, [Fe/H]=\,$-$1.52$\pm$0.08
and an age of 14.8$\pm$0.6\,Gyr, typical for the old
globulars in the Milky Way. Dolphin (\cite{dol00}) concluded that
the WLM started to form stars about 12\,Gyr ago, with
approximately half of the star-formation occurring during the
last 9\,Gyr, also based on {\it HST} imaging.

Recently, Venn et al. (\cite{ven03}) determined
[Fe/H]=\,$-0.38 \pm$0.29 from high-resolution spectroscopy of
two WLM blue supergiants. They found depleted stellar Oxygen abundance
by a factor of five, in comparison with the Oxygen abundance of
the H{\sc ii} regions. Even though later Lee et al.
(\cite{lee05}) reduced the discrepancy, some unusual chemical
evolution history is required to explain the observations because
the enhanced stellar abundance would put WLM well above the
metallicity-luminosity relation.

These results make it important to study the Carbon star in WLM
because of their sensitivity to the metal abundance. Besides, AGB
stars are representative for the stellar population with ages
between 1 -- 10\,Gyr and can be used as a observational
constraint to the properties of the post-main sequence stellar
evolution. Initially, the behavior of the C/M ratio is understood
at least qualitatively (Scalo \& Miller \cite{sca81}, Iben \& Renzini
\cite{ibe83}). Large number of studies of AGB stars in Local group galaxies
was performed using optical narrow band imaging. It is turn out to be
an easy way to separate M-type from Carbon stars (Albert et al. \cite{al00},
Battinelli \& Demers \cite{bat04}). The progress in the recent years
made it possible to produce the first observational calibrations of the
C/M ratio versus metallicity relation (Groenewegen \cite{gro06},
Cioni \& Habing \cite{cio05}, Battinelli \& Demers \cite{bat05}).

As AGB stars are almost the brightest cool stars it is easy to be
investigated in the near-IR range. The interest in this region of the
spectrum grow up strongly in the last three years (Cioni \& Habing \cite{cio03},
\cite{cio05}, Kang et al. \cite{kang05}, \cite{kang06}, Sohn et al.
\cite{sohn06}). Here we report analysis of the Carbon stars in WLM from near-IR
imaging.

\section{Observations and data reduction\label{SecData}}

The observations of WLM were made in Dec 2004 under
non-photometric conditions. In order to estimate the foreground
contamination an additional field located 14.8 arcmin South from the
center of the galaxy, far enough to eliminate the galaxy contribution,
was observed in Jul 2006 (hereafter "clear sky" field).
The near-IR imaged
and spectrograph SofI (Son of ISAAC) at the ESO NTT (New Technology
Telescope) on La Silla was used. The instrument is equipped with
1024$\times$1024 Hawaii HgCdTe array with 18.5 micron square pixels
and scale of 0.288 arcsec pixel$^{-1}$, giving a field of view of
4.92$\times$4.92 arcmin$^{2}$. We alternated between the target and a
clear patch of sky nearby. To minimize the effects of array
cosmetics and to improve the sky subtraction we introduced a
small random jitter of up to a few tens of arcseconds between
the sequential images. In total, we collected 10 images for both
fields in $J_S$ and 20 on the galaxy and 15 on the "clear sky"
field in $K_S$. Each of the images in $J_S$ is the average of
$2\times30$ s exposures and of $8\times8$ s exposures in $K_S$. The
total integration time was 10\,min in $J_S$ (hereafter $J$ for
simplicity) and in $K_S$ 21.33\,min on the galaxy and 16\,min on the
"clear sky".

\begin{figure}[ht]
\resizebox{\hsize}{!}{\includegraphics{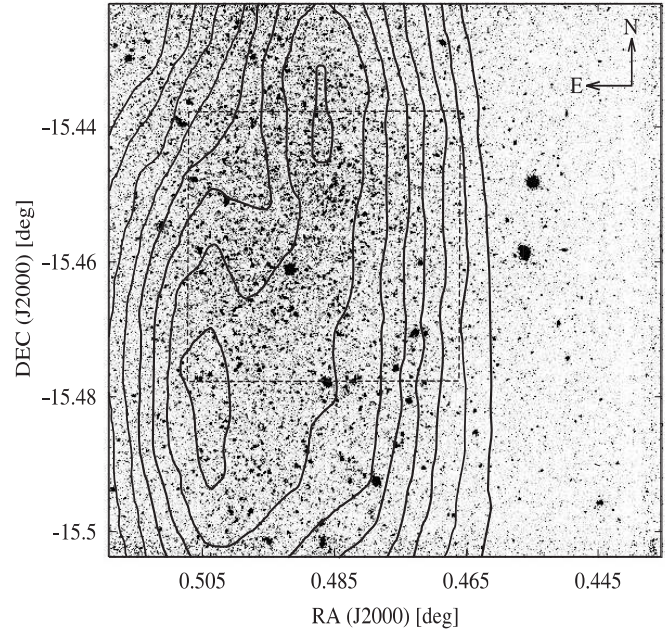}}
\caption{Near-IR image of WLM. Isolines show the H{\sc i}
map of Jackson et al. (\cite{jac04}) at levels from 25 to
95\,\% of the maximum column density (2.16$\times$10$^{21}$
cm$^{-2}$), in increments of 10\,\%. The square marks the
boundary of the "inner" field (see Section~\ref{SecData}).}
\label{WLM_final}
\end{figure}

The data reduction followed the typical steps. First, we
created a sky for each filter by median combination of the
off-target images using suitable upper limits and rejection
algorithms to exclude the stars present in the sky field. Next,
we subtracted the sky from each image, and divided it by the
dome flat. The dome flats are preferable in comparison with the
sky flats because they allow us to remove the variable bias
level, found in some Hawaii detectors. To account for the
uneven illumination of the dome flat screen, we applied an
illumination correction. It was derived from the photometry of
a standard star, placed on 16 positions across the field of average
view, in a 4$\times$4 grid. Finally, the individual images were
aligned and combined to form the final image
(Fig.~\ref{WLM_final}).

The photometric calibration was performed by comparing our
instrumental photometry with the measurements from the 2MASS
Point Source Catalog:
\begin{equation}
J = j - 5.55(\pm0.05)
\end{equation}
\begin{equation}
K_S = k - 8.65(\pm0.18)
\end{equation}
based on 18 common stars.
Here $J$ and $K_S$ are the 2MASS
systematic magnitudes, and $j$ and $k$ are the instrumental
magnitudes. No statistically significant color terms were
found.

\begin{figure}[!ht]
\resizebox{\hsize}{!}{\includegraphics{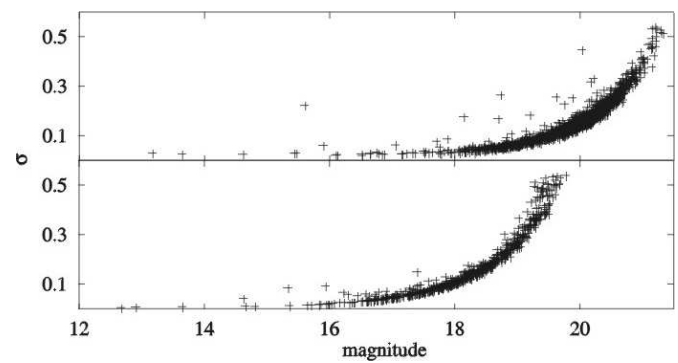}}
\caption{Photometric errors as a function of the magnitude for
$J$ (top) and $K_S$ (bottom) band. The completeness limits of
our photometry are $\sim$20.1 and $\sim$17.9\,mag for $J$ and
$K_S$-band,
respectively.}
\label{Fig_sigmas}
\end{figure}

The photometry was performed with the IRAF DAOPHOT package.
In galaxy field 1550 stars were detected in $J$ and 739 in $K_S$
with threshold of respectively 4$\sigma$ and  3$\sigma$ above
the background. The instrumental magnitudes for each filter
were obtained by using constant Gaussian PSF.
The positions and magnitudes of all measured stars matched
in both $J$ and $K_S$ frames with maximum tolerance of 3
pixels (0.86 arcsec) are given in Tabl.~\ref{Table_phot}. The
measurement errors are plotted in Fig.~\ref{Fig_sigmas}.

The reduction and calibration
procedures for "clear sky" field follow the analogical steps.
The photometric equations obtained from 7 2MASS stars
\begin{equation}
J = j - 3.12(\pm0.03)
\end{equation}
\begin{equation}
K_S = k - 5.34(\pm0.04)
\end{equation}
were used to transform the magnitudes of 89 stars in $J$ and 67 in
$K_S$ into 2MASS magnitudes. Matching procedure gives in total 50
common stars in $J$ and $K_S$.

\begin{table*}[t]
\begin{center}
\caption{Photometry of 555 WLM stars with $J$ and $K_S$-band
photometry.
The columns contain: identification number, RA and DEC (J2000),
$J$ and $K_S$ magnitudes (the 1$\sigma$ errors are given in
brackets.
Keys to the notes:
V - variable star, with identifier from the source paper
  (SC85 for Sandage \& Carlson \cite{san85} and
   M05 for Mora et al. \cite{mor05}),
C - Carbon star candidate (BD04 for Battinelli \& Demers \cite{bat04}),
G - globular cluster.
The paper edition of the journal contains only the first 20 entries. 
The rest of the table is available in the electronic version
of the journal only.}
\label{Table_phot}
\begin{tabular}{c@{   }c@{   }c@{   }c@{   }c@{   }c}
\hline
\multicolumn{1}{c}{No} &
\multicolumn{1}{c}{R.A.} &
\multicolumn{1}{c}{Dec.} &
\multicolumn{1}{c}{$J$} &
\multicolumn{1}{c}{$K_S$} &
\multicolumn{1}{c}{Note} \\
\hline
1 & 00:01:45.62 & $-$15:30:06.6 & 19.713 (0.135) & 18.268 (0.131) & M05\,LC\,120  \\
2 & 00:01:46.86 & $-$15:29:44.0 & 18.150 (0.047) & 16.737 (0.058) & C  \\
3 & 00:01:47.40 & $-$15:29:49.1 & 19.762 (0.119) & 18.376 (0.155) & C  \\
4 & 00:01:47.50 & $-$15:27:54.8 & 19.222 (0.073) & 18.402 (0.140) &   \\
5 & 00:01:47.99 & $-$15:27:47.5 & 19.408 (0.092) & 17.487 (0.065) & BD04 \\
6 & 00:01:48.20 & $-$15:26:49.0 & 19.327 (0.088) & 17.855 (0.097) & C  \\
7 & 00:01:48.34 & $-$15:28:08.5 & 19.402 (0.107) & 17.462 (0.096) & C  \\
8 & 00:01:48.42 & $-$15:27:08.6 & 19.078 (0.069) & 18.273 (0.118) &   \\
9 & 00:01:48.79 & $-$15:27:35.2 & 19.800 (0.130) & 18.849 (0.232) &   \\
10 & 00:01:48.80 & $-$15:27:16.8 & 19.132 (0.068) & 18.037 (0.108) & BD04 \\
11 & 00:01:49.09 & $-$15:27:06.3 & 19.712 (0.109) & 18.985 (0.255) &   \\
12 & 00:01:49.11 & $-$15:29:03.3 & 19.708 (0.114) & 19.501 (0.418) &   \\
13 & 00:01:49.26 & $-$15:26:53.4 & 13.177 (0.029) & 12.683 (0.004) &   \\
14 & 00:01:49.33 & $-$15:28:24.1 & 19.495 (0.092) & 18.920 (0.253) &   \\
15 & 00:01:49.50 & $-$15:26:06.4 & 18.821 (0.056) & 18.021 (0.103) &   \\
16 & 00:01:49.56 & $-$15:27:31.0 & 15.615 (0.222) & 14.631 (0.042) & G \\
17 & 00:01:49.57 & $-$15:28:00.6 & 18.921 (0.075) & 17.781 (0.090) &   \\
18 & 00:01:49.79 & $-$15:29:42.1 & 19.211 (0.076) & 18.764 (0.217) &   \\
19 & 00:01:49.88 & $-$15:26:32.3 & 18.833 (0.091) & 17.404 (0.090) & C  \\
20 & 00:01:49.94 & $-$15:26:20.7 & 19.196 (0.076) & 18.640 (0.192) &   \\
\hline
\end{tabular}
\end{center}
\end{table*}

\begin{figure}[!h]
\resizebox{\hsize}{!}{\includegraphics{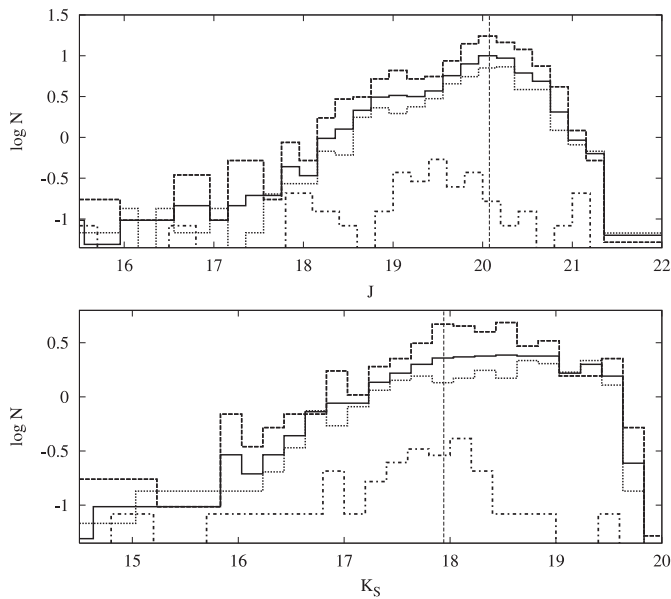}}
\caption{Luminosity functions for $J$ (upper panel) and $K_S$
(bottom panel) with a magnitude bin of 0.2 mag.
Log N is the number of stars normalized for 1 square arcmin.
Each panel shows the luminosity functions for the whole field
(solid lines), the inner part (dashed lines) and outer part
(dotted lines). The LF for the "clear sky" field is also
shown with dot-dashed lines. Vertical lines represent our
galaxy 100 \% photometry completeness for every band.}
\label{WLM_LF}
\end{figure}

The luminosity functions in $J$ and $K_S$ are shown in
Fig~\ref{WLM_LF}. The number of stars in each bin is
normalized to 1 square arcmin. We tested the completeness
limits variations across the galaxy by splitting the field
into an ``inner'' part, containing the area with maximum
H{\sc i} column densities and an ``outer'' part,
respectively 5.76 and 14.77 square arcmin (see
Fig~\ref{WLM_final}). The crowding effects are limited to
$\leq0.2$\,mag. We adopted $J^{\sc lim}$$\sim$20.1\,mag and
$K_S^{\sc lim}$$\sim$17.9\,mag.
For comparison the luminosity function of "clear sky" is also
shown in Fig~\ref{WLM_LF}. On average the LF is nearly an order
lower then the galaxy one.

\section{Color-Magnitude Diagram\label{Sec_CMD}}

\begin{figure*}[!ht]
\begin{center}
\resizebox{110mm}{!}{\includegraphics{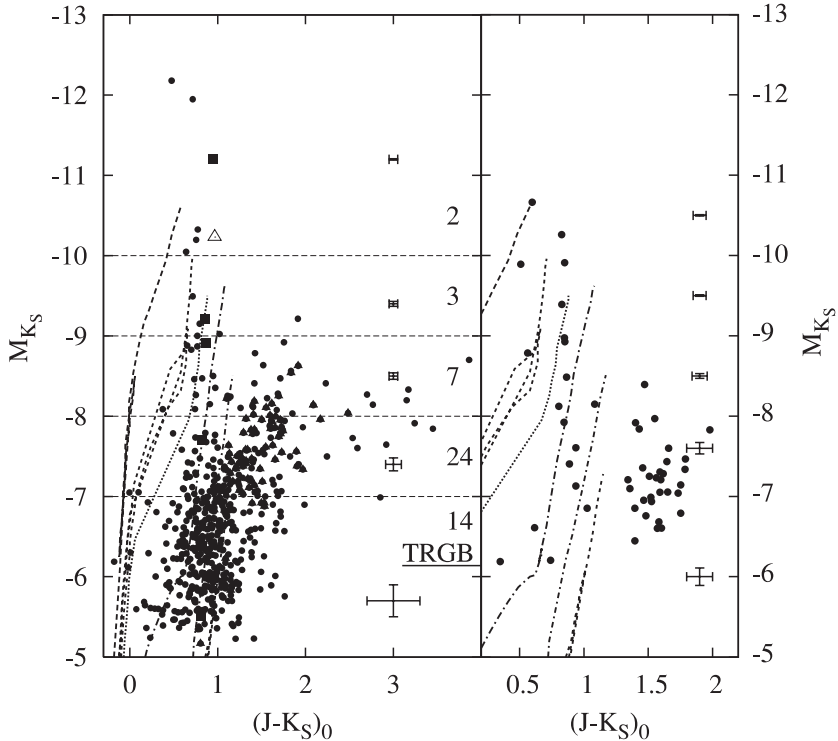}}
\caption{Near-IR $M_{K_S}$ versus ($J$ $-$ $K_S$)$_0$\,mag
color-magnitude diagram of WLM (left panel) and the
"clear sky" field (right panel; see Sec.~\ref{Sec_CMD} for
details).
The typical measurement errors for stars with
($J$ $-$ $K_S$)$_0$=1\,mag are shown on the right.
Isochrones for the 10, 20, 30, 100\,Myr, 1 and 10\,Gyr and
Z=0.001 from Bertelli et al. (\cite{ber94}) are plotted.
Left panel: the open triangle indicates the location of the
WLM globular cluster.
Six variable stars from Sandage \& Carlson (\cite{san85})
are plotted with solid squares and 77 Carbon stars of
Battinelli \& Demers (\cite{bat04}) are shown with solid
triangles. The level of the RGB tip is also marked in the
right.The number of foreground stars in every
1-magnitude bin along $M_{K_S}$ obtained from CMD of "clear sky"
are given on the right.}
\label{WLM_CMD}
\end{center}
\end{figure*}

The color-magnitude diagram of WLM is shown in
Fig.~\ref{WLM_CMD} (top panel). We used distance modulus
(m$-$M)$_0$=24.85\,mag -- an average of the estimates of:
(i) Minniti \& Zijlstra (\cite{min97}) who derived
(m$-$M)$_0$=24.75$\pm$0.1\,mag from the $I$-magnitude of
the red giant branch (hereafter RGB) tip and
(ii) Rejkuba et al. (\cite{rej00}) who obtained
(m$-$M)$_0$=24.95$\pm$0.13\,mag from $V$-band photometry of
the horizontal branch. These two values agree within the
errors and they are derived from different methods and data
sets.
Corrections for the Galactic extinction were made using
E($B$$-$$V$)=0.037\,mag (Schlegel et al. \cite{schleg98}).
Here and throughout the paper we use the extinction law of
Rieke \& Lebofsky (\cite{rieke85}).

For the high Galactic latitude of WLM (b= --$73.62^{\circ}$) we
expect the foreground contamination to be negligible. This is
confirmed from the number of stars found under similar
observational conditions in our "clear sky", which is just 9\% of
all stars in the galaxy field. Color-magnitude diagram was
construct using the same distance modulus and Galactic extinction
and is shown in Fig.~\ref{WLM_CMD} (right panel). In such way a
true estimation of the Galaxy contamination can be made by
counting the stars in 1 magnitude bins in $M_{K_S}$. The numbers
are given in the right on the Fig.~\ref{WLM_CMD} (left panel).

The comparison with isochrones of Bertelli et al.
(\cite{ber94}) indicate that the majority of the stars
are 1-10\,Gyr old, with a significant population of AGB
and Carbon stars with ($J$$-$$K_S$)$_0$$>$1\,mag. Only a
handful of blue stars with ($J$$-$$K_S$)$_0$$<$1\,mag and
$M_{K_S}$$>$$-$8\,mag are detected. We also identified
on our images a number of variable stars from Sandage \&
Carlson (\cite{san85}) and Mora et al. (\cite{mor05}).

The RGB properties versus metallicity calibrations of
Ivanov \& Borissova (\cite{iva02}) predict that the RGB
tip for the WLM metallicity is located at
$M_{K_S}$$>$$-$6.14\,mag. Unfortunately, the incompleteness
of the photometry in this magnitude range prevents us from
analysis of the RGB properties.

\section{Carbon stars identification and C/M
ratio in the WLM\label{Sec_Carbon_LF}}

WLM is unusually rich of Carbon star, along with two other
dwarf galaxies -- IC\,1613 and NGC\,6822 -- according to
Cook et al. (\cite{cook86}). Recently, Battinelli \& Demers
(\cite{bat04}) surveyed WLM using narrow band CN-TiO filters
and identified 149 Carbon stars and practically no M-type
AGB stars (77 identified Carbon stars in
our field are marked in Tabl.~\ref{Table_phot} with BD04).
They also found an extremely high C/M ratio of WLM:
12.4$\pm$3.7.

\begin{figure}[!h]
\resizebox{85mm}{!}{\includegraphics{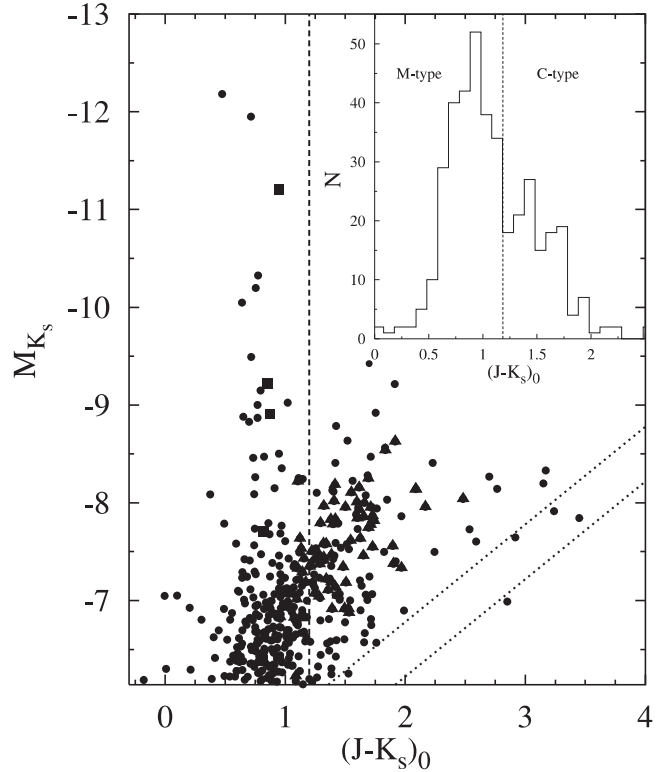}}
\caption{Color-magnitude $M_{K_S}$ versus $(J-K_S)_0$ diagram
for AGB stars in WLM. The two dotted lines represent the
100\,\% and 25\,\% photometry completeness limits. Six
variables from Sandage \& Carlson (\cite{san85}) are plotted
with solid squares and 77 Carbon stars from Battinelli \&
Demers (\cite{bat04}) are plotted with solid triangles.
The inset is a color distribution of the stars above the RGB
tip, used to separate the Oxygen rich and Carbon rich stars.
See Sec.~\ref{Sec_Carbon_LF} for details. The separation
line (dashed line) $(J-K_S)_0$=1.20 is shown on both panels.}
\label{CMD_AGB}
\end{figure}

The IR photometry allows us to carry out an independent census of
the Carbon stars, following the method applied on the Magellanic
Clouds (Cioni \& Habing \cite{cio03}), NGC\,6822 (Cioni \& Habing
\cite{cio05}, Kang et al. \cite{kang06}), NGC\,185 (Kang et al.
\cite{kang05}) and NGC\,147 (Sohn et al. \cite{sohn06}). It is
found that the color distribution of all AGB stars have a well
pronounced discontinuity at some point, M-type stars have a peak
follow by a red tail of Carbon stars. We separated the Carbon rich
from the Oxygen rich stars using the color distribution of the
stars above the RGB tip (Fig.~\ref{CMD_AGB}, the inset) and found
color limit at ($J$$-$$K_S$)$_0$=1.20\,mag. A similar dereddened
limit was found by Cioni \& Habing (\cite{cio05}) for NGC\,6822
and Sohn et al. (\cite{sohn06}) for NGC\,147:
($J$$-$$K_S$)$_0$=1.24 and ($J$$-$$K_S$)$_0$=1.28, respectively.

\begin{figure}[!h]
\resizebox{85mm}{!}{\includegraphics{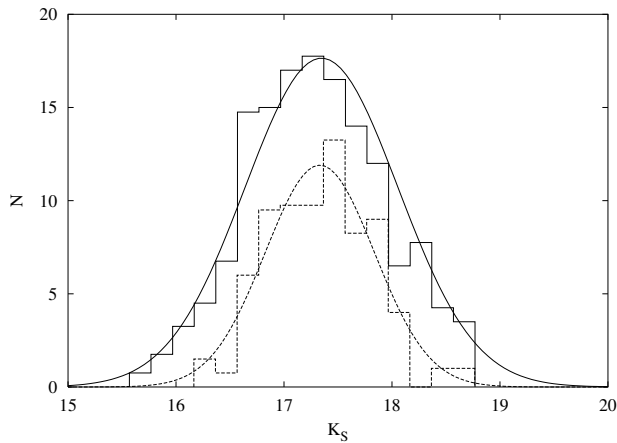}}
\caption{Luminosity functions of the Carbon stars in WLM.
The solid line represents the distribution of Carbon stars
above the RGB tip. The dashed line shows the
77 Carbon stars from Battinelli \& Demers (\cite{bat04})
that we identified in our field. The bin width is 0.2\,mag.
The final Gaussian fits to the data are shown (see
Sec.~\ref{Sec_Carbon_LF} for details).}
\label{LF_Carbon}
\end{figure}

Integrating over the two parts of the histogram we obtain
146 Carbon stars and 259 M stars. The latter number includes
all stars with spectral types M0 and later (M0, M1, M2 etc.)
which sometimes is designated as M0+. Using the same criteria on
the ``clear sky'' field we obtain 31 C stars and 19 M stars
as all of them are above the RGB tip. Taking into account the
latter numbers this will reduce our C stars to 115 and
the M star to 240.

The $K_S$ luminosity functions for our Carbon star sample and for
the 77 cross-identified Carbon stars (shown in Fig.~\ref{CMD_AGB}
with solid triangles) from Battinelli \& Demers (\cite{bat04}) are
shown in Fig.~\ref{LF_Carbon}. To eliminate the binning effects we
created for each data set two luminosity functions shifting the
bin centers by half of the bin width and averaged them. We fitted
the luminosity function with a Gaussian, in two iterations,
excluding the values that deviated more than 3$\sigma$. The final
fits have mean of $<$$K_S$$>$=17.35\,mag, $\sigma$=0.71\,mag for
our sample and $<$$K_S$$>$=17.33\,mag, $\sigma$=0.51\,mag for the
77 Carbon stars from Battinelli \& Demers (\cite{bat04}) which
corresponds to $< M_{K_S}>$=$-7.51$\,mag and
$<$$M_{K_S}$$>$=$-$7.53\,mag, respectively. 

Previous studies in
some LG galaxies in the near IR lead to identical mean absolute K
magnitude of the C stars: $<$$M_{K_S}$$>$=$-$7.93$\pm$0.36\,mag
(Kang et al. \cite{kang05}), $<$$M_{K_S}$$>$=$-$7.56$\pm$0.47\,mag
(Sohn et al. \cite{sohn06}) and
$<$$M_{K_S}$$>$=$-$7.60$\pm$0.50\,mag (Kang et al. \cite{kang06})
for NGC\,185, NGC\,147 and NGC\,6822, respectively.
$<$$M_{K_S}$$>$ in WLM is comparable with these values in 1$\sigma$ 
level and support this assumption. The sample of
galaxies include dwarf irregulars and dwarf ellipticals but is not
representative for wide range of metallicity and a possible 
influence over the mean absolute K magnitude can not be
entirely excluded.

Next, we derived a C/M ratio of 0.56$\pm$0.12 (foreground
corrected ratio 0.48), using the Carbon and M-type stars we
identified above the RGB tip. However, Cioni \& Habing
(\cite{cio05}) pointed that this classification method omits the
early type Carbon stars that might be bluer than the adopted color
limit. The incompleteness due to this effect can be estimated by
comparison of our Carbon star list and that of Battinelli \&
Demers (\cite{bat04}) -- it yielded additional 15 Carbon stars
among the M stars, bringing the C/M ratio up to 0.66$\pm$0.11
(0.58). The difference between the two values is small,
considering the Poisson uncertainties given here.

Both values obtained here are smaller by a factor of 20 than the
result of Battinelli \& Demers (\cite{bat04}): C/M0+$\sim$12.4.
The large difference is due to the presence of M-type AGB
population, which is almost entirely absent in Battinelli \&
Demers (\cite{bat04}) survey of WLM. 
We can only speculate that this difference is related to the narrow band filter 
centered at TiO absorption feature -- prominent for spectral types later than M3. 
Their method probably omits early M type stars. Furthermore, the low metallicity of WLM 
diminishes additionally the TiO feature strenght and makes the detection of M type 
stars more difficult.

\section{Spatial distribution of the stellar populations}

\begin{figure*}[!ht]
\resizebox{175mm}{!}{\includegraphics{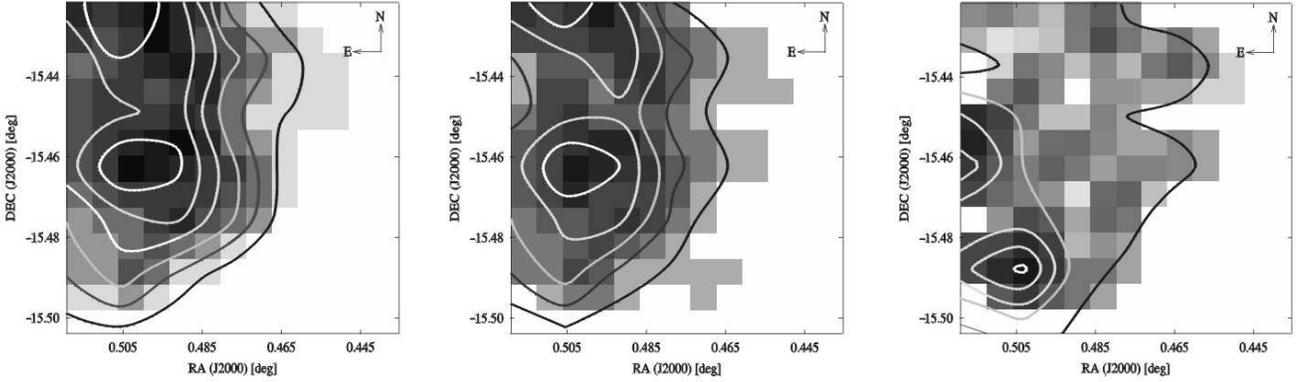}}
\caption{Spatial distribution of all stars (left), AGB
stars (middle) and C/M ratio (right panel). The isolines
are at levels: 1 to 6 with a step of 1, 1 to 5 with a step
of 1, and 0.2 to 1.4 with a step of 0.3, respectively for
the three plots from left to right.
}
\label{Spatial_plot}
\end{figure*}

Our data cover most of the main body of WLM, allowing us to
study the spatial distribution of stellar populations. We
created 2-dimensional histograms (Fig.~\ref{Spatial_plot})
with 22.5$\times$22.5\,arcsec$^{2}$ bins, corresponding to linear
scale of 100$\times$100\,pc at the distance of the galaxy.
The distributions were smoothed by a box car function of
width 2.

The histograms indicate no major difference between the
stellar distributions. There are two peaks, corresponding
to complexes with typical sizes of a few hundred parsecs.
The C/M ratio exhibits a more complicated behavior --
the peak is slightly offset from the peak of the stellar
distributions. Further studies are necessary to verify if
this reflects a real metallicity or an age gradient
across the galaxy.

\section{Miscellaneous}

\subsection{Cepheid distance to WLM in the near-IR
\label{Sec_Cepheids}}

We identified 5 Cepheids from the list of Sandage \& Carlson
(\cite{san85}) on our $J$-band image (the $K_S$ image is too
shallow to detect them). We derived a distance to WLM adopting
the period-luminosity relation of Gieren, Fouque \& Gomez
(\cite{gie98}):
\begin{equation}
M_J = -5.240(\pm0.028)-3.129(\pm0.052)\times(log P - 1.0)
\label{PL_relation}
\end{equation}
where $M_J$ is the absolute $J$-band magnitude of the Cepheids
and P is the period in days.

\begin{figure}[!h]
\resizebox{85mm}{!}{\includegraphics{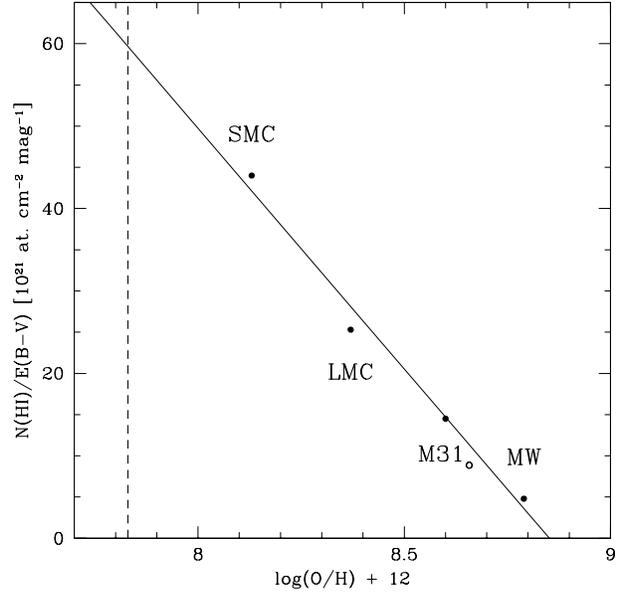}}
\caption{Gas-to-dust ratio versus log(O/H)+12 relation.
Data points are labeled (see Sec.~\ref{Sec_Cepheids} for
details). The solid line is the linear least squares fit,
and the vertical dashed line indicates the WLM abundance
from Lee et al. (\cite{lee05}).}
\label{G-t-D_ratio_fit}
\end{figure}

The distances to individual Cepheids are listed in
Tabl.~\ref{Table_Ceph}. We removed the Milky Way extinction
using $A_J$=0.033\,mag. To correct for the internal extinction
towards individual Cepheids we used the H{\sc i} map of
Jackson et al. (\cite{jac04}). The hydrogen surface densities
were converted into extinction assuming gas-to-dust ratio
N(H{\sc i})/E($B$$-$$V$)=60$\pm$10.0~[10$^{21}$ at. cm$^{-2}$
mag$^{-1}$]. This value is an extrapolation based on a linear
least squares fit (Fig.~\ref{G-t-D_ratio_fit}) of the
gas-to-dust ratio versus log(O/H)+12:
\begin{equation}
N(H{\sc i})/E(B-V)=-(58\pm5)\times[log(O/H)+12]+(517\pm44)
\end{equation}
with r.m.s.=2.

This relation is based on gas-to-dust ratio estimates for:
the solar neighborhood (Bohlin et al. \cite{bohl78}; marked
MW in Fig.~\ref{G-t-D_ratio_fit}), the M\,31 outskirts
(Cuillandre et al. \cite{cul01}; M\,31$_{o}$), LMC (Junkkarinen
et al. \cite{junk04}; LMC) and SMC (Jakobsson et al.
\cite{jak04}; SMC). The abundances are taken from Russell
\& Dopita (\cite{rus90}) for LMC and SMC. In case of the
Milky Way we used the Galactic metallicity gradient from
Esteban et al. (\cite{est05}), and for M\,31 -- the metallicity
gradient of Smartt et al. (\cite{sma01}). Note that this is a
preliminary calibration, based on a limited number of
measurements. More gas-to-dust ratios are necessary to
derive a reliable relation.

Finally, we assumed that the Cepheids are located on the WLM
mid-plane and therefore, the extinctions are due only to half
of the total hydrogen column densities. The extinction
intrinsic to WLM does not exceed 0.015\,mag, due to the high
gas-to-dust ratio of the galaxy.

Rejecting the outlier No.\,66 (see Table~\ref{Table_Ceph}), we
obtain a median distance modulus of
(m$-$M)$_0$=24.84$\pm$0.14\,mag. We tentatively assume that the
error is given by the maximum deviation. In addition, there
are systematic errors of 0.03\,mag for the zero-point of the
period-luminosity relation (Equ.~\ref{PL_relation}), and
0.02\,mag for the LMC distance adopted in Gieren et al.
(\cite{gie98}).

\begin{table*}[!ht]
\begin{center}
\caption{Cepheid distance to WLM.
The columns contain: identification number, $J$-band magnitude
(the 1$\sigma$ errors are given in brackets), period in days
from Sandage \& Carlson (\cite{san85}), absolute magnitude
calculates from Equ.~\ref{PL_relation}, apparent distance
modulus, total extinction in $A_J$ derived from gas-to-dust
ratio (including Milky Way extinction of $A_J$=0.033 mag,
Schlegel et al. \cite{schleg98}) and true distance
modulus (m$-$M)$_0$. }
\label{Table_Ceph}
\begin{tabular}{c@{}c@{}c@{}c@{}c@{}c@{}c}
\hline
\multicolumn{1}{c}{No} &
\multicolumn{1}{c}{$J$}\hspace{0.0cm} &
\multicolumn{1}{c}{Period, days}\hspace{0.0cm} &
\multicolumn{1}{c}{$M_J$}\hspace{0.0cm} &
\multicolumn{1}{c}{(m$-$M)$_J$}\hspace{0.0cm} &
\multicolumn{1}{c}{$A_J$}\hspace{0.0cm} &
\multicolumn{1}{c}{(m$-$M)$_0$}\hspace{0.0cm} \\
\hline
 7 & 20.108 (0.155) & 7.6712  & -4.880 & 24.988 & 0.048 & 24.940\\
 8 & 20.016 (0.150) & 7.3754  & -4.827 & 24.843 & 0.047 & 24.796\\
21 & 19.829 (0.124) & 9.33306 & -5.146 & 24.975 & 0.044 & 24.931\\
48 & 19.724 (0.160) & 8.53000 & -5.024 & 24.748 & 0.048 & 24.700\\
66 & 20.197 (0.189) & 4.54714 & -4.170 & 24.367 & 0.040 & 24.327\\
\hline
\end{tabular}
\end{center}
\end{table*}

\subsection{Variable stars in WLM}

Some variable stars from the lists of Sandage \& Carlson
(\cite{san85}) and Mora et al. (in preparation) were included in
our field. They are marked in Tabl.~\ref{Table_phot}.

\subsection{Structural parameters of WLM\label{Sec_struct}}

\begin{figure}[!h]
\resizebox{\hsize}{!}{\includegraphics{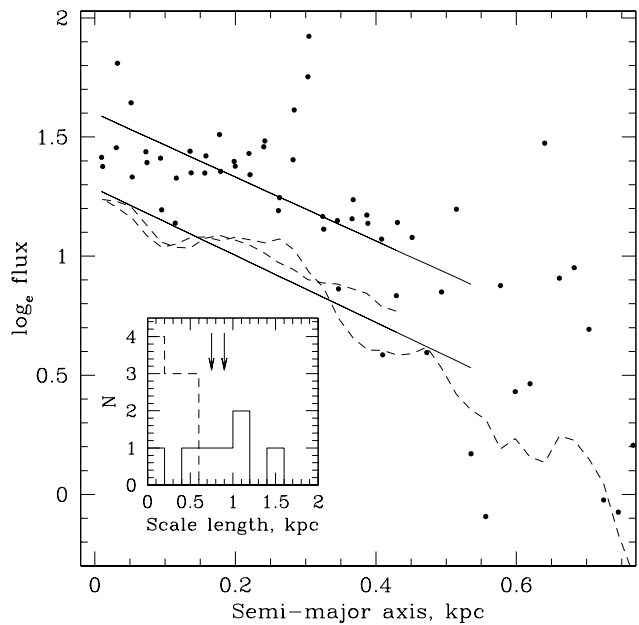}}
\caption{Scale length of the WLM disk in $J$-band. The dots
indicates the average values along the semi-major axis, the
dashed line is the same for the median-averaged image. The
solid lines are the fits.
The inset shows the distribution of the scale lengths for the
dwarf irregulars (solid line) and dwarf spheroidal galaxies
(dashed lines) in the Local group. The left arrow indicates
our estimate and the right arrow shows estimate of Fisher
\& Tully (\cite{fish75}).}
\label{Fig_scale}
\end{figure}

Structural parameters of galaxies are determined more reliably
from IR images rather than optical ones because the effects of
the intrinsic extinction are diminished. Our images do not
encompass the entire galaxy but nevertheless we are able to
determine approximately the scale length along the major axis.
To do that we averaged the pixel values in a 63.4-arcsec wide
strip ($\sim$300\,pc at the distance of WLM), crossing the
center of the galaxy in North-South direction. We used the
$J$-band image because of the higher signal-to-noise ratio. The
average flux values along the semi-major axis are shown in
Fig.~\ref{Fig_scale}. An exponential law fit over the central
1\,kpc region of WLM yielded a scale length of
0.75$\pm$0.14\,kpc, similar to the scale length of 0.9\,kpc
determined by Fisher \& Tully (\cite{fish75}). The inset in
Fig.~\ref{Fig_scale} shows histograms of the scale lengths for
dwarf galaxies in the Local Group from the compilation of
Mateo (\cite{mat98}).

The bright -- and younger -- stars in the galaxy are resolved
on our images, but the fainter -- and older ones -- not.
We addressed the question if there is a different scale length
for different stellar populations by excluding the bright
stars from this analysis: we averaged the $J$-band image with
a 61$\times$61 pixel median filter, effectively removing the
resolved stars. The derived scale length for ``older''
population was 0.71$\pm$0.04\,kpc, again for the central
1\,kpc of the galaxy. Our estimates have to be treated with
caution because the strip is not perfectly aligned along the
major axis.

\section{Discussion and summary}

This work adds another object to the list of galaxies with
homogeneously determined C/M0+ ratios from near-IR photometry:
LMC, SMC, NGC\,6822, NGC\,147, NGC\,185 and now the WLM. This is
an opportunity for independent check of the results from
narrow band Carbon star surveys. The number of data points is
still limited and did not allow us to test if using only C/M
ratios derived from IR observations can reduce the spread of the
relation but this is the objective of our further studies. We
obtained for the WLM a C/M ratio of $\sim$0.6$\pm$0.1 and a peak
of the Carbon stars luminosity function of
$<$$M_{K_S}$$>$=$-$7.51\,mag.

To measure the intrinsic reddening in WLM we used the H{\sc i}
map of WLM (Jackson et al. (\cite{jac04}), and we adopted a
linear dependence of the gas-to-dust ratio on the metallicity:
N(H{\sc i})/E($B$$-$$V$)=$-$(58$\pm$5)$\times$[log(O/H)+12]+(517$\pm$44). 
The gas-to-dust ratio for the WLM is 60$\pm$10~[10$^{21}$
at. cm$^{-2}$ mag$^{-1}$]. This allows us to measure the 
extinction along the line of sight towards individual Cepheids 
and by period-luminosity relation for the Cepheids (Gieren 
\cite{gie98}) in the IR we determined the distance to 
the WLM -- (m$-$M)$_0$=24.84$\pm$0.14\,mag, similar to the 
previous estimates. To probe the significance of used PL 
relation we re-determined the distance modulus using the 
PL relation calibrated by Testa et al. (\cite{testa06}) for the 
Cepheids in two LMC young clusters. This results in distance 
modulus between 24.76 and 25.00, depending on the assumed 
pulsation models. 
Finally, we fitted the surface
brightness distribution along the main axis of WLM with an
exponential low yielding a scale length of 0.75$\pm$0.14\,kpc,
typical value for the dwarf irregular galaxies.

\vspace{0.5cm}
\begin{acknowledgements}
This work obtained a limited support by the grants
F-1302/03 and VU-F-201/06 of the Bulgarian Scientific Foundation.
Also, this publication makes use of data products from the
Two Micron All Sky Survey, which is a joint project of the
University of Massachusetts and the IR Processing and
Analysis Center/California Institute of Technology, funded
by the National Aeronautics and Space Administration and
the National Science Foundation.
This research has made use of the SIMBAD database, operated
at CDS, Strasbourg, France.
The authors thank Dr. M.-R. Cioni for the useful comments and 
the anonymous referee for the help in improving the article.

\end{acknowledgements}

\end{document}